\documentclass[aps,pra,twocolumn,showpacs,groupedaddress]{revtex4}
\usepackage{graphicx} 
\usepackage{amssymb}
\usepackage{amsmath}

\begin{document}

\title{Positron attachment to the He doubly excited states}  
\author{M. W. J. Bromley}
\affiliation{Centre for Quantum-Atom Optics, School of Mathematics and Physics, The University of Queensland, Brisbane QLD 4075, Australia}
\author{J. Mitroy}
\affiliation{Centre for Antimatter-Matter Studies and School of Engineering, Charles Darwin University, Darwin NT 0909, Australia}
\author{K. Varga}
\affiliation{Department of Physics and Astronomy, Vanderbilt University, Nashville, Tennessee 37235, USA} 

\date{\today}

\begin{abstract}
The projection method is used to demonstrate the existence of positron 
attachment to three doubly excited states of helium.  The $e^+$He($2s^2$ $^1$S$^e$),
$e^+$He($3s^2$ $^1$S$^e$), and the $e^+$He($2s2p$ $^3$P$^o$) states have
binding energies of 0.447 eV, 0.256 eV and 0.486 eV respectively.
These energies were computed with the stochastic variational method
and the configuration interaction method. 
These states will exist as resonances in the $e^+$-He continuum and 
the $e^+$He($2s^2$ $^1$S$^e$) state could be detectable in the 
$e^+$+He collision spectrum.  
A resonance width of $0.068$ eV was computed for the $e^+$He($2s^2$ $^1$S$^e$),
state by using the complex rotation method.   The existence of a series of 
$e^+$He($ns^2$ $^1$S$^e$) resonances associated with the He($ns^2$) 
double Rydberg series is also predicted and an explicit calculation 
demonstrating the existence of the e$^+$He($3s^2$ $^1$S$^e$) state is 
reported.  
 
\end{abstract}


\maketitle 

In this letter the ability of a positron to attach itself to the 
doubly excited states of helium is demonstrated by explicit calculation 
using the Feshbach projection operator approach that was used in some of 
the earliest calculations of the helium doubly excited spectrum    
\cite{bhatia67a,bhatia75a}. Besides the intrinsic interest in such  
exotic Coulomb systems, the result provides a pathway to providing
experimental confirmation that positrons can be attached to 
electrically neutral atoms to form bound states.  

It is now widely accepted that positrons can form bound states with a
variety of atoms \cite{ryzhikh98e,mitroy02b,schrader10a}.
While the evidence for positron binding is strong, it is
derived from calculation.  Binding energies range from $0.0129$ eV in
the case of $e^+$Na \cite{mitroy05d} to $0.50$ eV for the $e^+$Ca
ground state \cite{bromley06c}.

There is solid experimental evidence that positrons can form bound states
with a variety of molecules.  The energy resolved positron annihilation
cross sections for a number of molecules (e.g. C$_3$H$_8$, C$_6$H$_{14}$)
show features that have been identified as Feshbach resonances formed by
the trapping of positrons in vibrationally excited states of molecules \cite{barnes03a}.
This is thought to be the mechanism responsible for the large positron
annihilation rates observed for many molecules in gas-phase positron
annihilation spectroscopy experiments \cite{gribakin10a}.

While the experimental evidence of positron binding to molecules is good,
there is no experimental evidence that could be construed as
demonstrating the existence of positron-atom bound states.  One possible
signature would be the existence of resonant structures associated with
atomic excited states in the positron scattering spectrum.
Years of experimentation, however, have revealed little evidence for
the existence of resonant states in positron-atom scattering spectra
\cite{sullivan01b,mitroy02b,surko05a}.

A number of schemes have been put forward to demonstrate the existence 
of positron-atom bound states \cite{mitroy99e,mitroy01e,bromley02d,dzuba10b}.   
The most recent proposal suggested that positron scattering experiments
be performed on open shell transition-metal atoms having polarizabilities and
ionization energies conducive to binding positrons \cite{dzuba10b}.
Open shell systems are recommended since such
systems would have low-lying excited states that could also
bind a positron.  Positron binding to low-lying excited
states would result in Feshbach resonances appearing in the low-energy
annihilation cross section.  However, the transition-metal atoms
most likely to bind a positron represent difficult
propositions for experimentation.

The present letter demonstrates that three of the doubly excited 
states of helium, namely the He($2s^2$ $^1$S$^e$), He($3s^2$ $^1$S$^e$) 
and He($2s2p$ $^3$P$^o$) states can attach a positron 
with attachment energies exceeding 0.250 eV.  The 
$e^+$He($2s^2$ $^1$S$^e$) and $e^+$He($3s^2$ $^1$S$^e$) states
manifest themselves as resonances in the $e^+$+He continuum.
A positron cannot excite the He($2s2p$ $^3$P$^o$) state from the 
He($1s^2$ $^1$S$^e$) ground state since there is no 
exchange interaction between the positron and electrons.
These states can be regarded as analogues the triply excited 
negative ion resonances seen in the electron-helium
spectrum at $57$-$61$ eV incident energy
\cite{kuyatt65a,quemener71a,hicks74a,vanderburgt86a,trantham99a}.
%
\begin{table}[th]
\caption[] { \label{HeMg}
Energies (in a.u.) of some He states given with respect to the 
He$^{2+}$ threshold.  Three sets of helium energies are given.  
One set, $E_{\rm CR}$ are taken from complex rotation calculations, 
the two other sets are taken from projection operator calculations.  The 
projection operator energies in the $E_{\rm QHQ}$ column come from calculations 
that use a Hylleraas basis, while those in the $E_{\rm CI}$ column come
from CI calculations as described in the text.  There is no complex rotation 
energy for the He($2p^2$ $^3$P$^e$) state since it is a bound state.
}  
\begin{ruledtabular}
\begin{tabular}{lccclc}
State   &  \multicolumn{1}{c}{$E_{\rm CR}$}  &  $E_{\rm QHQ}$ & $E_{\rm CI}$  \\ \hline 
He$^+$($2s$)         &  $-$0.500000              & $-$0.500000                  & $-$0.500000  \\       
He($2s^2$ $^1$S$^e$) &  $-$0.777818 \cite{ho86a} & $-$0.778774 \cite{bhatia67a} & $-$0.778781  \\       
He($2s2p$ $^3$P$^o$) &  $-$0.760498 \cite{ho93b} & $-$0.761492 \cite{bhatia75a} & $-$0.761492  \\
He($2p^2$ $^3$P$^e$) &                           & $-$0.710500 \cite{bhatia70a} & $-$0.710500  \\
He($2p^2$ $^1$D$^e$) &  $-$0.701946 \cite{ho91b} & $-$0.702817 \cite{bhatia72a} & $-$0.702819   \\
He($2s2p$ $^1$P$^o$) &  $-$0.69314 \cite{ho81a}  & $-$0.692895 \cite{bhatia69a} & $-$0.692897  \\
He($2p^2$ $^1$S$^e$) &  $-$0.621928 \cite{ho86a} & $-$0.622744 \cite{bhatia75a} & $-$0.622736    \\       
\end{tabular}
\end{ruledtabular}
\end{table}

\begin{figure}[th]
\centering{
\includegraphics[width=8.25cm,angle=0]{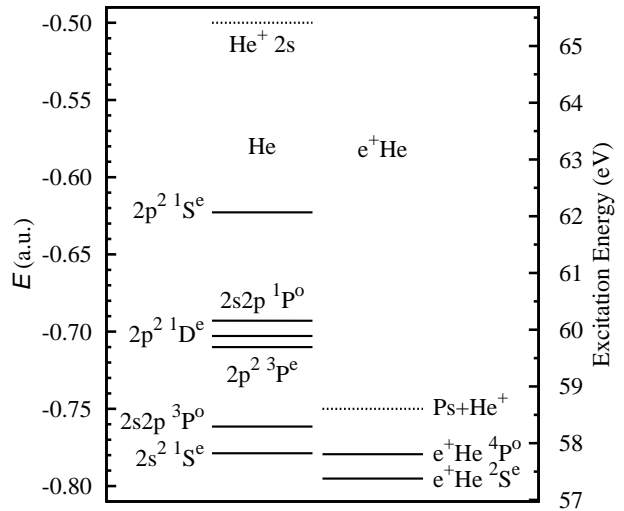}
}
\caption[]{ \label{fig:helevels}
Energy level diagram showing the positions of the He doubly excited 
states and the states with an attached positron.  The position of 
the Ps($1s$)+He$^+$($2s$) threshold is also shown.  The axis on the
right gives the positron collision energy (in eV) needed to excite 
these states. 
}
\end{figure}

One motivation for the present investigation was the realization that 
the doubly excited states of helium have energetics very similar 
to those of the Mg atom which binds a positron with a binding energy 
of 0.465 eV \cite{mitroy08e} and also supports a prominent $p$-wave 
shape resonance in the elastic scattering channel at 0.096 eV 
incident energy \cite{mitroy08e,savage11a}.  The binding energy of the 
Mg$^+$($3s$) ground state is $-$0.55254 a.u. \cite{nistasd315} 
while the He$^+$($2s$) binding energy is $-$0.50 a.u.. The binding 
energy of the Mg($3s^2$) ground state with respect to the 
Mg$^+$($3s$) threshold is  $-$0.2810 a.u., while the binding 
energy of the He($2s^2$) resonance with respect to the 
He$^+$($2s$) state is $-$0.2778 a.u.  \cite{ho86a}.  The respective 
dipole polarizabilities, calculated with oscillator strength 
sum rules \cite{mitroy10a}, are 76.2 $a_0^3$ for He($2s^2$) and 
71.3 $a_0^3$ for Mg($3s^2$) \cite{mitroy10a}.   
The He energies are listed in Table \ref{HeMg} 
and plotted in Fig.~(\ref{fig:helevels}).

The projection method \cite{bhatia67a,bhatia75a} provides a 
computational strategy for the identification of resonances.
In this method, the electrons are not allowed to occupy those 
low-lying states that could result in the auto-ionization of the 
system.  The projection method energies, $E_{\rm QHQ}$, of the 
helium doubly excited states in Table \ref{HeMg}, computed
using Hylleraas basis sets, differ from those determined by the
dynamically complete complex rotation method by less
than 0.001 a.u..  The projection method has successfully
been applied to calculate the positions of the He$^-$ resonances
associated with the He doubly excited states \cite{bylicki92a}.
Here, the Hamiltonian was chosen for the $N=2$ electron and
one positron system to be
\begin{eqnarray} 
{\hat H} &=& -\sum_{i=1}^{N+1} \frac{\nabla_i^{2}}{2} 
 - \sum_{i=1}^N \frac{2}{r_i}  + \frac{2}{r_{N+1}} \nonumber \\ 
  &+& \sum_{i<j}^N \frac{1}{|{\mathbf r}_{i}-{\mathbf r}_j|}   
  - \sum_{i=1}^N \frac{1}{|{\mathbf r}_{N+1}-{\mathbf r}_i|} .
\label{scatHam}
\end{eqnarray} 
Investigation of resonant states requires diagonalizing the 
Hamiltonian ${\hat Q}{\hat H}{\hat Q}$, where the
projection operator ${\hat Q} = (1 - {\hat P})$.   
For the $n=2$ helium doubly excited states one can use
combinations of the single particle projection operator
${\hat P}_i = |\phi_{1s}({\bf r}_i)\rangle \langle \phi_{1s}({\bf r}_i)| \equiv |1s\rangle \langle 1s|$
where $\phi_{1s}({\bf r}_i)$ is the wavefunction of the
He$^+$($1s$) orbital \cite{bylicki92a}.

Two independent computational methods, the configuration interaction (CI) 
and the stochastic variational method (SVM) \cite{varga95,ryzhikh98e},   
are used to diagonalize ${\hat Q}{\hat H}{\hat Q}$.
The He$^+$($1s$) ground state is excluded from the CI wavefunction
by Schmidt-orthogonalizing the $\ell = 0$ single-particle electron
orbital basis to the He$^+$($1s$) state.  This obviates the need for 
the inclusion of an explicit projection operator since 
$(\langle 1s | \otimes \langle n\ell|) | \Psi \rangle = 0$ will automatically
be satisfied by the CI basis that is used to diagonalize Eq.~(\ref{scatHam}).
The single particle $|n\ell \rangle$ in the present calculations were
chosen to be Laguerre type orbitals (LTOs). 

The CI method was initially applied to the calculation of the He doubly 
excited states.  
The basis included 49 LTOs for $\ell=0$,
and 50 LTOs for the other $\ell$'s.  The largest $\ell$ value used
in these calculations was $\ell = 8$.  The CI energies are given
in Table~\ref{HeMg}, and were extrapolated to the $\ell = \infty$ limit
using a procedure described shortly.
They agree with the $E_{\rm QHQ}$ energies to within $10^{-5}$ a.u..

The $e^+$He CI basis was constructed by letting the two electrons
and the positron form all of the possible configurations with
a fixed electron-electron spin ($S_e$),
total angular momentum ($L_T$), and total wavefunction parity ($\pi$),
subject to the further selection rules,
$\max(\ell_0,\ell_1,\ell_2) \le J$, and
$\min(\ell_1,\ell_2) \le L_{\rm int}$, and
$(-1)^{(\ell_0+\ell_1+\ell_2)} = -1^{\pi}$.
%
%
In these rules $\ell_0$, $\ell_1$ and $\ell_2$ are respectively
the orbital angular momenta of the positron and the two electrons,
with a maximum single-particle orbital angular momentum of $J$.
The number of LTOs for each $\ell$ was $15$ with the exception of
$\ell = 0,1,2$, and $3$ where $18,18,17$, and $16$ LTOs were used.
The parameter $L_{\rm int}$ was set to 4.
The largest $\ell$ in the orbital space was $J=12$ for the
$e^+$He($2s^2$ $^1$S$^e$) state and $J = 9$ for the
$e^+$He($2s2p$ $^3$P$^o$) state.

The main technical problem afflicting CI calculations of positron-atom 
interactions is the slow convergence of the energy with $J$
\cite{mitroy02b,mitroy06a}.  One way to
determine the $J \rightarrow \infty$ energy, $\langle E \rangle_{\infty}$,
is to use an asymptotic analysis.  It has been shown that successive
increments, $\Delta E_{J} = \langle E \rangle_J - \langle E \rangle_{J-1}$,
to the energy can written as \cite{hill85a,mitroy06a,bromley07a}:
\begin{equation}
\Delta E_J \approx \frac {A_E}{(J+{\scriptstyle \frac{1}{2}})^4}
    + \frac {B_E}{(J+{\scriptstyle \frac{1}{2}})^5}
    + \frac {C_E}{(J+{\scriptstyle \frac{1}{2}})^6} .
\label{extrap1}
\end{equation}
The $J \to \infty$ limit, is determined by fitting sets of
$\langle E \rangle_J$ values to Eq.~(\ref{extrap1}).
The coefficients, $A_E$, $B_E$ and $C_E$  
are determined at a particular $J$ from 4 successive energies 
($\langle E\rangle_{J-3}$, $\langle E\rangle_{J-2}$,
$\langle E \rangle_{J-1}$ and $\langle E \rangle_{J}$).
Once the coefficients have been determined it is easy to obtain the
$J \to \infty$ limit.   Application of asymptotic series analysis
to helium has resulted in CI calculations reproducing the ground
state energy to an accuracy of $10^{-8}$ a.u. \cite{bromley07a}.  

The CI energy of the $e^+$He($2s^2$) state (see Table \ref{tab:positronHe}) 
was $-$0.795058 a.u.. Subtracting this from the He($2s^2$) $E_{\rm CI}$ 
of $-$0.778781 a.u. gives a binding energy of 0.016277 a.u..  This binding 
energy is an underestimate since the energy of the He($2s^2$) state in 
the CI basis used for the $e^+$He calculation was $-$0.778771 a.u..
The $J \to \infty$ extrapolation contributed 10$\%$  
to the binding energy.  The $e^+$He($2s^2$) 
binding energy is only 4$\%$ smaller than the binding energy of the 
positron to the Mg($3s^2$) ground state, 
namely 0.01704 a.u. \cite{mitroy08e}.

The He($2s2p$ $^3$P$^o$) state also binds a positron with a binding
energy of $0.017870$ a.u..  The surprisingly large binding energy is caused by 
the relatively small excitation energy of $0.051$ a.u. from 
the He($2s2p$ $^3$P$^o$) state to the He($2p^2$ $^3$P$^e$).  This leads to  
the He($2s2p$ $^3$P$^o$) state having a static dipole polarizability of 
$157$ $a_0^3$. The $e^+$He($2s2p$ $^3$P$^o$) positron binding energy is 
larger than that of the $e^+$Be($2s2p$ $^3$P$^o$) state which is only 
$0.000087$ a.u. \cite{bromley07b}. 

The SVM was also used to determine the energy of the resonance state using the
projection ansatz.   The SVM diagonalizes the Hamiltonian in a basis of
explicitly correlated gaussians (ECGs).  The non-linear parameters of the
ECG basis are optimized by a trial and error process.  Such a process is
possible since the ECG matrix elements of the Hamiltonian are
very easy to compute.  The diagonalization of ${\hat Q}{\hat H}{\hat Q}$
is approximated by adding an orthogonalizing
pseudo-projector (OPP) \cite{krasnopolsky74,ryzhikh98e,mitroy99h},
to the Hamiltonian to exclude the He$^+$($1s$) state from
being occupied.  The modified Hamiltonian is
\begin{equation}
{\hat H}_{\rm OPP} = {\hat H} + \lambda {\hat P}_{\rm OPP}
\end{equation}
where $\lambda$ is chosen to be a large positive number.
The operator ${\hat P}_{\rm OPP}$ is defined as 
\begin{equation}
{\hat P}_{\rm OPP} = |\phi_{1s}({\bf r}_1)\rangle \langle \phi_{1s}({\bf r}_1)| 
   + |\phi_{1s}({\bf r}_2)\rangle \langle \phi_{1s}({\bf r}_2)| 
\end{equation}
where $\phi_{1s}({\bf r}_i)$ again refers to the He$^+$($1s$) state.
Any part of the wavefunction with a non-zero overlap
with the He$^+$($1s$) state tends to increase the energy.  
The energy minimization inherent to the SVM leads to a   
ground state wavefunction with a very small overlap with the
He$^+$($1s$) state.  The parameter $\lambda$ was set to $10^6$ a.u. for
the present calculations.  The He$^+$($1s$) state was expanded as a
linear combination of 12 gaussians.  

\begin{table}[th]
\caption[] { \label{tab:positronHe}
Calculated energies of some $e^+$He states.  The CI calculations are
also given with a $J \to \infty$ correction as discussed in the text.
The binding energies are denoted by $\varepsilon$.
 }  
\begin{ruledtabular}
\begin{tabular}{llcccc}
State   &  Method & $J$ & $\langle E \rangle_J$ (a.u.) & $\varepsilon$ (a.u.) & $\varepsilon$ (eV) \\ \hline 
$e^+$He($2s^2$ $^1$S$^e$) & CI  & 12       &  $-$0.793537 &  0.014756 & 0.4015 \\       
                          & CI  & $\infty$ &  $-$0.795058 &  0.016277 & 0.4429 \\       
                          & SVM &  ---     &  $-$0.795210 &  0.016429 & 0.4471 \\       
$e^+$He($2s2p$ $^3$P$^o$) & CI  & 9        &  $-$0.776306 &  0.014814 & 0.4031 \\
                          & CI  & $\infty$ &  $-$0.779362 &  0.017869 & 0.4863 \\
$e^+$He($3s^2$ $^1$S$^e$) & CI  & 12       &  $-$0.468860 &   --      & -- \\
                          & CI  & $\infty$ &  $-$0.481643 &  0.009420 & 0.2563 \\
\end{tabular}
\end{ruledtabular}
\end{table}

The SVM energy of the He$(2s^2)$ state
was $-0.778786$ a.u., i.e. $5 \times 10^{-6}$ a.u. below the CI energy.
The dimension of the largest SVM calculation of
the $e^+$He$(2s^2)$ state was 900 ECGs.  The SVM binding energy of the
positron to the He$(2s^2)$ state given in Table \ref{tab:positronHe} was
$0.016429$ a.u.  Examination of the convergence pattern suggests that
the SVM energy is within $2 \times 10^{-4}$ a.u. of the variational limit.
The SVM and CI binding energies for this state are in
excellent agreement when the respective uncertainties arising from
finite size basis sets are taken into consideration.

The $e^+$He($2s^{2} \ ^{1}S^e$) system is also likely to support 
a $^2$P$^o$ shape resonance just above the He($2s^{2} \ ^{1}S^e$) 
threshold.  This is based on the similarity of the He and Mg
polarizabilities and the positron attachment energies in the
$^2$S$^e$ channel.   The $e^+$-Mg $^2$P$^o$ shape resonance 
was located at $0.00351$ a.u. above the elastic scattering threshold 
and had a width of $0.00396$ a.u. \cite{mitroy08e}.    

It is likely that there will be an infinite series of resonances 
associated with the set of He($ns^2$) doubly excited states.  An 
investigation of the ($m^{2+}$, 2$e^-$, $e^+$) system revealed that 
this system remains bound when the mass $m^{2+} \to 0$ \cite{mitroy02f}.   
Decreasing the $m^{2+}$ mass weakens the effective strength of the 
$m^{2+}$-$e^-$ interaction and provides an analogue of the  
He$^{2+}$-$ns(e^{-})$ interaction.  A first test was performed by a CI  
investigation of the $e^+$He(3$\ell$,3$\ell'$)  systems.  In this
case the single particle basis was orthogonalized to the He$^+$($1s,2s,2p$) 
states.   The CI energy of the He($3s^2$) state is $-0.354562$ a.u.  
Since the removal energy of the electron with respect to the He$^+$($3\ell$) 
threshold, $-0.132340$ a.u., is less than the positronium ground state energy of 
$-$0.25 a.u., the threshold for attaching a positron to the He($3s^2$) 
state is at $-0.47222\bar{2}$ a.u..

The CI calculation for the $e^+$He(3$s^2$)  state gave an energy of 
$-$0.481643 a.u. The binding energy of this state is 0.009420 a.u.  
The stability of this system provides strong evidence for an  
infinite number of $e^+$He($ns^2$) type resonances.  It is 
likely that the rich resonance structures of the PsH system 
\cite{drachman79b} will be replicated for positron interactions 
with the doubly excited helium atoms. 
    
Reference can be made to $e^{-}$+He scattering experiments to give a 
first order estimate on the viability of experimental detection.  
A number of electron scattering experiments have 
demonstrated electron attachment to the He doubly excited states 
\cite{kuyatt65a,quemener71a,hicks74a,vanderburgt86a,trantham99a}.  
Experiments that detect total cross sections involving ground state 
atoms and ions probably do not have a sufficiently large signal to 
background ratio to detect the $e^+$He resonances.  For example, 
He$^+$ ions were detected in the experiment of 
Quemener {\em et al.} \cite{quemener71a}.  There, the 
cross section for the creation of He$^+$ varied by only $1\%$ over the 
width of the He$^-$($2s^22p$) resonance.     
Higher signal to background ratios have been achieved in $e^-$-He experiments 
that measured differential cross sections \cite{hicks74a,vanderburgt86a}.

Finally, the widths of the resonances and energy resolution of positron beams 
need to be considered.  Modern trap-based positron beams can achieve 
a total energy resolution of about $40$ meV \cite{gribakin10a,sullivan08a}. 
An indication of the resonance widths can be made by reference to
the widths of their doubly excited parent states.  The width of the He($2s^2$) 
state is $\Gamma = 123$ meV \cite{ho81a} and the 
He($2s2p \ ^{3}$P$^o$) state is $\Gamma = 8.1$ meV \cite{ho81a}.  
The widths of He$^-$ resonances based on these parents, that of 
the He$^-(2s^22p \ ^{2}$P$^o)$ state of $71$ meV and that of the 
He$^-(2s2p^{2} \ ^{4}$P$^o)$ state of $10.3$ meV \cite{bylicki95a} 
are reflective of their two electron parents.  We performed an
SVM complex rotation calculation \cite{ho83a}
by augmenting the ECG basis with additional functions representing
the $e^+$+He and He$^+$+Ps continuum.
The energy shifted to $-0.79484$ a.u. and the width was $0.00249$ a.u.
($68$ meV), which is large enough to detect with current positron beam 
technology.  Previously known positron-atom resonances are either 
too narrow as in the case of hydrogen and sodium \cite{zhou95b,han08b},    
or involve atoms which do not naturally exist in gaseous form 
\cite{han08b,mitroy08e,dzuba10b}.

JM was supported under the Australian Research Council's
(ARC) Centre of Excellence program.  MB was supported by
an ARC Future Fellowship (FT100100905), and NSF grants PHY-0970127
and CHE-0947087.  The authors thank Mr. Joshua Machacek, Prof. Stephen Buckman
and Dr. James Sullivan for the suggestion that we investigate
the doubly excited helium systems.


\begin{thebibliography}{46}
\expandafter\ifx\csname natexlab\endcsname\relax\def\natexlab#1{#1}\fi
\expandafter\ifx\csname bibnamefont\endcsname\relax
  \def\bibnamefont#1{#1}\fi
\expandafter\ifx\csname bibfnamefont\endcsname\relax
  \def\bibfnamefont#1{#1}\fi
\expandafter\ifx\csname citenamefont\endcsname\relax
  \def\citenamefont#1{#1}\fi
\expandafter\ifx\csname url\endcsname\relax
  \def\url#1{\texttt{#1}}\fi
\expandafter\ifx\csname urlprefix\endcsname\relax\def\urlprefix{URL }\fi
\providecommand{\bibinfo}[2]{#2}
\providecommand{\eprint}[2][]{\url{#2}}

\bibitem[{\citenamefont{Bhatia et~al.}(1967)\citenamefont{Bhatia, Temkin, and
  Perkins}}]{bhatia67a}
\bibinfo{author}{\bibfnamefont{A.~K.} \bibnamefont{Bhatia}},
  \bibinfo{author}{\bibfnamefont{A.}~\bibnamefont{Temkin}}, \bibnamefont{and}
  \bibinfo{author}{\bibfnamefont{J.~F.} \bibnamefont{Perkins}},
  \bibinfo{journal}{Phys. Rev.} \textbf{\bibinfo{volume}{153}},
  \bibinfo{pages}{177} (\bibinfo{year}{1967}).

\bibitem[{\citenamefont{{Bhatia} and {Temkin}}(1975)}]{bhatia75a}
\bibinfo{author}{\bibfnamefont{A.~K.} \bibnamefont{{Bhatia}}} \bibnamefont{and}
  \bibinfo{author}{\bibfnamefont{A.}~\bibnamefont{{Temkin}}},
  \bibinfo{journal}{\pra} \textbf{\bibinfo{volume}{11}}, \bibinfo{pages}{2018}
  (\bibinfo{year}{1975}).

\bibitem[{\citenamefont{Ryzhikh et~al.}(1998)\citenamefont{Ryzhikh, Mitroy, and
  Varga}}]{ryzhikh98e}
\bibinfo{author}{\bibfnamefont{G.~G.} \bibnamefont{Ryzhikh}},
  \bibinfo{author}{\bibfnamefont{J.}~\bibnamefont{Mitroy}}, \bibnamefont{and}
  \bibinfo{author}{\bibfnamefont{K.}~\bibnamefont{Varga}},
  \bibinfo{journal}{J.~Phys.~B} \textbf{\bibinfo{volume}{31}},
  \bibinfo{pages}{3965} (\bibinfo{year}{1998}).

\bibitem[{\citenamefont{Mitroy et~al.}(2002)\citenamefont{Mitroy, Bromley, and
  Ryzhikh}}]{mitroy02b}
\bibinfo{author}{\bibfnamefont{J.}~\bibnamefont{Mitroy}},
  \bibinfo{author}{\bibfnamefont{M.~W.~J.} \bibnamefont{Bromley}},
  \bibnamefont{and} \bibinfo{author}{\bibfnamefont{G.~G.}
  \bibnamefont{Ryzhikh}}, \bibinfo{journal}{J.~Phys.~B}
  \textbf{\bibinfo{volume}{35}}, \bibinfo{pages}{R81} (\bibinfo{year}{2002}).

\bibitem[{\citenamefont{Schrader}(2010)}]{schrader10a}
\bibinfo{author}{\bibfnamefont{D.~M.} \bibnamefont{Schrader}}, in
  \emph{\bibinfo{booktitle}{Proceedings of the International School of Physics
  ``Enrico Fermi'', 7-17 July 2009}}, edited by
  \bibinfo{editor}{\bibfnamefont{A.}~\bibnamefont{Dupasquier}}
  \bibnamefont{and} \bibinfo{editor}{\bibfnamefont{A.~P.}
  \bibnamefont{Mills~Jr.}} (\bibinfo{publisher}{Italiana di Fisica},
  \bibinfo{address}{Italy}, \bibinfo{year}{2010}), p. \bibinfo{pages}{337}.

\bibitem[{\citenamefont{Mitroy}(2005)}]{mitroy05d}
\bibinfo{author}{\bibfnamefont{J.}~\bibnamefont{Mitroy}},
  \bibinfo{journal}{Phys.~Rev.~Lett.} \textbf{\bibinfo{volume}{94}},
  \bibinfo{pages}{033402} (\bibinfo{year}{2005}).

\bibitem[{\citenamefont{Bromley and Mitroy}(2006)}]{bromley06c}
\bibinfo{author}{\bibfnamefont{M.~W.~J.} \bibnamefont{Bromley}}
  \bibnamefont{and} \bibinfo{author}{\bibfnamefont{J.}~\bibnamefont{Mitroy}},
  \bibinfo{journal}{Phys.~Rev.~A} \textbf{\bibinfo{volume}{73}},
  \bibinfo{pages}{032507} (\bibinfo{year}{2006}).

\bibitem[{\citenamefont{Barnes et~al.}(2003)\citenamefont{Barnes, Gilbert, and
  Surko}}]{barnes03a}
\bibinfo{author}{\bibfnamefont{L.~D.} \bibnamefont{Barnes}},
  \bibinfo{author}{\bibfnamefont{S.~J.} \bibnamefont{Gilbert}},
  \bibnamefont{and} \bibinfo{author}{\bibfnamefont{C.~M.} \bibnamefont{Surko}},
  \bibinfo{journal}{Phys.~Rev.~A} \textbf{\bibinfo{volume}{67}},
  \bibinfo{pages}{032706} (\bibinfo{year}{2003}).

\bibitem[{\citenamefont{{Gribakin} et~al.}(2010)\citenamefont{{Gribakin},
  {Young}, and {Surko}}}]{gribakin10a}
\bibinfo{author}{\bibfnamefont{G.~F.} \bibnamefont{{Gribakin}}},
  \bibinfo{author}{\bibfnamefont{J.~A.} \bibnamefont{{Young}}},
  \bibnamefont{and} \bibinfo{author}{\bibfnamefont{C.~M.}
  \bibnamefont{{Surko}}}, \bibinfo{journal}{Rev.~Mod.~Phys.}
  \textbf{\bibinfo{volume}{82}}, \bibinfo{pages}{2557} (\bibinfo{year}{2010}).

\bibitem[{\citenamefont{Sullivan et~al.}(2001)\citenamefont{Sullivan, Gilbert,
  Buckman, and Surko}}]{sullivan01b}
\bibinfo{author}{\bibfnamefont{J.~P.} \bibnamefont{Sullivan}},
  \bibinfo{author}{\bibfnamefont{S.~J.} \bibnamefont{Gilbert}},
  \bibinfo{author}{\bibfnamefont{S.~J.} \bibnamefont{Buckman}},
  \bibnamefont{and} \bibinfo{author}{\bibfnamefont{C.~M.} \bibnamefont{Surko}},
  \bibinfo{journal}{J.~Phys.~B} \textbf{\bibinfo{volume}{34}},
  \bibinfo{pages}{L467} (\bibinfo{year}{2001}).

\bibitem[{\citenamefont{Surko et~al.}(2005)\citenamefont{Surko, Gribakin, and
  Buckman}}]{surko05a}
\bibinfo{author}{\bibfnamefont{C.~M.} \bibnamefont{Surko}},
  \bibinfo{author}{\bibfnamefont{G.~F.} \bibnamefont{Gribakin}},
  \bibnamefont{and} \bibinfo{author}{\bibfnamefont{S.~J.}
  \bibnamefont{Buckman}}, \bibinfo{journal}{J.~Phys.~B}
  \textbf{\bibinfo{volume}{38}}, \bibinfo{pages}{R57} (\bibinfo{year}{2005}).

\bibitem[{\citenamefont{Mitroy and Ryzhikh}(1999{\natexlab{a}})}]{mitroy99e}
\bibinfo{author}{\bibfnamefont{J.}~\bibnamefont{Mitroy}} \bibnamefont{and}
  \bibinfo{author}{\bibfnamefont{G.~G.} \bibnamefont{Ryzhikh}},
  \bibinfo{journal}{J.~Phys.~B} \textbf{\bibinfo{volume}{32}},
  \bibinfo{pages}{L411} (\bibinfo{year}{1999}{\natexlab{a}}).

\bibitem[{\citenamefont{Mitroy and Ivanov}(2001)}]{mitroy01e}
\bibinfo{author}{\bibfnamefont{J.}~\bibnamefont{Mitroy}} \bibnamefont{and}
  \bibinfo{author}{\bibfnamefont{I.~A.} \bibnamefont{Ivanov}},
  \bibinfo{journal}{J.~Phys.~B} \textbf{\bibinfo{volume}{34}},
  \bibinfo{pages}{L121} (\bibinfo{year}{2001}).

\bibitem[{\citenamefont{Bromley and Mitroy}(2002)}]{bromley02d}
\bibinfo{author}{\bibfnamefont{M.~W.~J.} \bibnamefont{Bromley}}
  \bibnamefont{and} \bibinfo{author}{\bibfnamefont{J.}~\bibnamefont{Mitroy}},
  \bibinfo{journal}{Phys.~Rev.~A} \textbf{\bibinfo{volume}{65}},
  \bibinfo{pages}{062506} (\bibinfo{year}{2002}).

\bibitem[{\citenamefont{{Dzuba} et~al.}(2010)\citenamefont{{Dzuba}, {Flambaum},
  and {Gribakin}}}]{dzuba10b}
\bibinfo{author}{\bibfnamefont{V.~A.} \bibnamefont{{Dzuba}}},
  \bibinfo{author}{\bibfnamefont{V.~V.} \bibnamefont{{Flambaum}}},
  \bibnamefont{and} \bibinfo{author}{\bibfnamefont{G.~F.}
  \bibnamefont{{Gribakin}}}, \bibinfo{journal}{Phys.~Rev.~Lett.}
  \textbf{\bibinfo{volume}{105}}, \bibinfo{pages}{203401}
  (\bibinfo{year}{2010}).

\bibitem[{\citenamefont{{Kuyatt} et~al.}(1965)\citenamefont{{Kuyatt},
  {Simpson}, and {Mielczarek}}}]{kuyatt65a}
\bibinfo{author}{\bibfnamefont{C.~E.} \bibnamefont{{Kuyatt}}},
  \bibinfo{author}{\bibfnamefont{J.~A.} \bibnamefont{{Simpson}}},
  \bibnamefont{and} \bibinfo{author}{\bibfnamefont{S.~R.}
  \bibnamefont{{Mielczarek}}}, \bibinfo{journal}{Phys.~Rev.}
  \textbf{\bibinfo{volume}{138}}, \bibinfo{pages}{385} (\bibinfo{year}{1965}).

\bibitem[{\citenamefont{Qu\'em\'ener et~al.}(1971)\citenamefont{Qu\'em\'ener,
  Paquet, and Marmet}}]{quemener71a}
\bibinfo{author}{\bibfnamefont{J.~J.} \bibnamefont{Qu\'em\'ener}},
  \bibinfo{author}{\bibfnamefont{C.}~\bibnamefont{Paquet}}, \bibnamefont{and}
  \bibinfo{author}{\bibfnamefont{P.}~\bibnamefont{Marmet}},
  \bibinfo{journal}{Phys. Rev. A} \textbf{\bibinfo{volume}{4}},
  \bibinfo{pages}{494} (\bibinfo{year}{1971}).

\bibitem[{\citenamefont{Hicks et~al.}(1974)\citenamefont{Hicks, Cvejanovic,
  Comer, and Read}}]{hicks74a}
\bibinfo{author}{\bibfnamefont{P.~J.} \bibnamefont{Hicks}},
  \bibinfo{author}{\bibfnamefont{S.}~\bibnamefont{Cvejanovic}},
  \bibinfo{author}{\bibfnamefont{J.}~\bibnamefont{Comer}}, \bibnamefont{and}
  \bibinfo{author}{\bibfnamefont{F.~H.} \bibnamefont{Read}},
  \bibinfo{journal}{Vacuum} \textbf{\bibinfo{volume}{24}}, \bibinfo{pages}{573}
  (\bibinfo{year}{1974}).

\bibitem[{\citenamefont{{van der Burgt} et~al.}(1986)\citenamefont{{van der
  Burgt}, {van Eck}, and {Heideman}}}]{vanderburgt86a}
\bibinfo{author}{\bibfnamefont{P.~J.~M.} \bibnamefont{{van der Burgt}}},
  \bibinfo{author}{\bibfnamefont{J.}~\bibnamefont{{van Eck}}},
  \bibnamefont{and} \bibinfo{author}{\bibfnamefont{H.~G.~M.}
  \bibnamefont{{Heideman}}}, \bibinfo{journal}{J.~Phys.~B}
  \textbf{\bibinfo{volume}{19}}, \bibinfo{pages}{2015} (\bibinfo{year}{1986}).

\bibitem[{\citenamefont{{Trantham} et~al.}(1999)\citenamefont{{Trantham},
  {Jacka}, {Rau}, and {Buckman}}}]{trantham99a}
\bibinfo{author}{\bibfnamefont{K.~W.} \bibnamefont{{Trantham}}},
  \bibinfo{author}{\bibfnamefont{M.}~\bibnamefont{{Jacka}}},
  \bibinfo{author}{\bibfnamefont{A.~R.~P.} \bibnamefont{{Rau}}},
  \bibnamefont{and} \bibinfo{author}{\bibfnamefont{S.~J.}
  \bibnamefont{{Buckman}}}, \bibinfo{journal}{J.~Phys.~B}
  \textbf{\bibinfo{volume}{32}}, \bibinfo{pages}{815} (\bibinfo{year}{1999}).

\bibitem[{\citenamefont{Ho}(1986)}]{ho86a}
\bibinfo{author}{\bibfnamefont{Y.~K.} \bibnamefont{Ho}},
  \bibinfo{journal}{Phys. Rev. A} \textbf{\bibinfo{volume}{34}},
  \bibinfo{pages}{4402} (\bibinfo{year}{1986}).

\bibitem[{\citenamefont{Ho}(1993)}]{ho93b}
\bibinfo{author}{\bibfnamefont{Y.~K.} \bibnamefont{Ho}},
  \bibinfo{journal}{Phys. Rev. A} \textbf{\bibinfo{volume}{48}},
  \bibinfo{pages}{3598} (\bibinfo{year}{1993}).

\bibitem[{\citenamefont{Bhatia}(1970)}]{bhatia70a}
\bibinfo{author}{\bibfnamefont{A.~K.} \bibnamefont{Bhatia}},
  \bibinfo{journal}{Phys. Rev. A} \textbf{\bibinfo{volume}{2}},
  \bibinfo{pages}{1667} (\bibinfo{year}{1970}).

\bibitem[{\citenamefont{Ho and Bhatia}(1991)}]{ho91b}
\bibinfo{author}{\bibfnamefont{Y.~K.} \bibnamefont{Ho}} \bibnamefont{and}
  \bibinfo{author}{\bibfnamefont{A.~K.} \bibnamefont{Bhatia}},
  \bibinfo{journal}{Phys. Rev. A} \textbf{\bibinfo{volume}{44}},
  \bibinfo{pages}{2895} (\bibinfo{year}{1991}).

\bibitem[{\citenamefont{Bhatia}(1972)}]{bhatia72a}
\bibinfo{author}{\bibfnamefont{A.~K.} \bibnamefont{Bhatia}},
  \bibinfo{journal}{Phys. Rev. A} \textbf{\bibinfo{volume}{6}},
  \bibinfo{pages}{120} (\bibinfo{year}{1972}).

\bibitem[{\citenamefont{Ho}(1981)}]{ho81a}
\bibinfo{author}{\bibfnamefont{Y.~K.} \bibnamefont{Ho}},
  \bibinfo{journal}{Phys. Rev. A} \textbf{\bibinfo{volume}{23}},
  \bibinfo{pages}{2137} (\bibinfo{year}{1981}).

\bibitem[{\citenamefont{Bhatia and Temkin}(1969)}]{bhatia69a}
\bibinfo{author}{\bibfnamefont{A.~K.} \bibnamefont{Bhatia}} \bibnamefont{and}
  \bibinfo{author}{\bibfnamefont{A.}~\bibnamefont{Temkin}},
  \bibinfo{journal}{Phys. Rev.} \textbf{\bibinfo{volume}{182}},
  \bibinfo{pages}{15} (\bibinfo{year}{1969}).

\bibitem[{\citenamefont{Mitroy et~al.}(2008)\citenamefont{Mitroy, Zhang,
  Bromley, and Young}}]{mitroy08e}
\bibinfo{author}{\bibfnamefont{J.}~\bibnamefont{Mitroy}},
  \bibinfo{author}{\bibfnamefont{J.~Y.} \bibnamefont{Zhang}},
  \bibinfo{author}{\bibfnamefont{M.~W.~J.} \bibnamefont{Bromley}},
  \bibnamefont{and} \bibinfo{author}{\bibfnamefont{S.~I.} \bibnamefont{Young}},
  \bibinfo{journal}{Phys.~Rev.~A} \textbf{\bibinfo{volume}{78}},
  \bibinfo{pages}{012715} (\bibinfo{year}{2008}).

\bibitem[{\citenamefont{{Savage} et~al.}(2011)\citenamefont{{Savage}, {Fursa},
  and {Bray}}}]{savage11a}
\bibinfo{author}{\bibfnamefont{J.~S.} \bibnamefont{{Savage}}},
  \bibinfo{author}{\bibfnamefont{D.~V.} \bibnamefont{{Fursa}}},
  \bibnamefont{and} \bibinfo{author}{\bibfnamefont{I.}~\bibnamefont{{Bray}}},
  \bibinfo{journal}{\pra} \textbf{\bibinfo{volume}{83}}, \bibinfo{eid}{062709}
  (\bibinfo{year}{2011}).

\bibitem[{\citenamefont{Ralchenko et~al.}(2008)\citenamefont{Ralchenko,
  Kramida, Reader, and {NIST ASD Team}}}]{nistasd315}
\bibinfo{author}{\bibfnamefont{Y.}~\bibnamefont{Ralchenko}},
  \bibinfo{author}{\bibfnamefont{A.}~\bibnamefont{Kramida}},
  \bibinfo{author}{\bibfnamefont{J.}~\bibnamefont{Reader}}, \bibnamefont{and}
  \bibinfo{author}{\bibnamefont{{NIST ASD Team}}}, \emph{\bibinfo{title}{{NIST
  Atomic Spectra Database Version 3.1.5}}} (\bibinfo{year}{2008}),
  \urlprefix\url{http://physics.nist.gov/asd3}.

\bibitem[{\citenamefont{Mitroy et~al.}(2010)\citenamefont{Mitroy, Safronova,
  and Clark}}]{mitroy10a}
\bibinfo{author}{\bibfnamefont{J.}~\bibnamefont{Mitroy}},
  \bibinfo{author}{\bibfnamefont{M.~S.} \bibnamefont{Safronova}},
  \bibnamefont{and} \bibinfo{author}{\bibfnamefont{C.~W.} \bibnamefont{Clark}},
  \bibinfo{journal}{J.~Phys.~B} \textbf{\bibinfo{volume}{43}},
  \bibinfo{pages}{202001} (\bibinfo{year}{2010}).

\bibitem[{\citenamefont{Bylicki}(1992)}]{bylicki92a}
\bibinfo{author}{\bibfnamefont{M.}~\bibnamefont{Bylicki}},
  \bibinfo{journal}{Phys. Rev. A} \textbf{\bibinfo{volume}{45}},
  \bibinfo{pages}{2079} (\bibinfo{year}{1992}).

\bibitem[{\citenamefont{Varga and Suzuki}(1995)}]{varga95}
\bibinfo{author}{\bibfnamefont{K.}~\bibnamefont{Varga}} \bibnamefont{and}
  \bibinfo{author}{\bibfnamefont{Y.}~\bibnamefont{Suzuki}},
  \bibinfo{journal}{Phys.~Rev.~C} \textbf{\bibinfo{volume}{52}},
  \bibinfo{pages}{2885} (\bibinfo{year}{1995}).

\bibitem[{\citenamefont{Mitroy and Bromley}(2006)}]{mitroy06a}
\bibinfo{author}{\bibfnamefont{J.}~\bibnamefont{Mitroy}} \bibnamefont{and}
  \bibinfo{author}{\bibfnamefont{M.~W.~J.} \bibnamefont{Bromley}},
  \bibinfo{journal}{Phys.~Rev.~A} \textbf{\bibinfo{volume}{73}},
  \bibinfo{pages}{052712} (\bibinfo{year}{2006}).

\bibitem[{\citenamefont{Hill}(1985)}]{hill85a}
\bibinfo{author}{\bibfnamefont{R.~N.} \bibnamefont{Hill}},
  \bibinfo{journal}{J.~Chem.~Phys.} \textbf{\bibinfo{volume}{83}},
  \bibinfo{pages}{1173} (\bibinfo{year}{1985}).

\bibitem[{\citenamefont{Bromley and Mitroy}(2007{\natexlab{a}})}]{bromley07a}
\bibinfo{author}{\bibfnamefont{M.~W.~J.} \bibnamefont{Bromley}}
  \bibnamefont{and} \bibinfo{author}{\bibfnamefont{J.}~\bibnamefont{Mitroy}},
  \bibinfo{journal}{Int.~J.~Quantum~Chem.} \textbf{\bibinfo{volume}{107}},
  \bibinfo{pages}{1150} (\bibinfo{year}{2007}{\natexlab{a}}).

\bibitem[{\citenamefont{Bromley and Mitroy}(2007{\natexlab{b}})}]{bromley07b}
\bibinfo{author}{\bibfnamefont{M.~W.~J.} \bibnamefont{Bromley}}
  \bibnamefont{and} \bibinfo{author}{\bibfnamefont{J.}~\bibnamefont{Mitroy}},
  \bibinfo{journal}{Phys.~Rev.~A} \textbf{\bibinfo{volume}{75}},
  \bibinfo{pages}{042506} (\bibinfo{year}{2007}{\natexlab{b}}).

\bibitem[{\citenamefont{{Krasnopol'sky} and Kukulin}(1974)}]{krasnopolsky74}
\bibinfo{author}{\bibfnamefont{V.~M.} \bibnamefont{{Krasnopol'sky}}}
  \bibnamefont{and} \bibinfo{author}{\bibfnamefont{V.~I.}
  \bibnamefont{Kukulin}}, \bibinfo{journal}{Sov.~J.~Nucl.~Phys.}
  \textbf{\bibinfo{volume}{20}}, \bibinfo{pages}{883} (\bibinfo{year}{1974}),
  \bibinfo{note}{yad.Fiz.(USSR) \textbf{20} (1974) 883}.

\bibitem[{\citenamefont{Mitroy and Ryzhikh}(1999{\natexlab{b}})}]{mitroy99h}
\bibinfo{author}{\bibfnamefont{J.}~\bibnamefont{Mitroy}} \bibnamefont{and}
  \bibinfo{author}{\bibfnamefont{G.~G.} \bibnamefont{Ryzhikh}},
  \bibinfo{journal}{Comput.~Phys.~Commun.} \textbf{\bibinfo{volume}{123}},
  \bibinfo{pages}{103} (\bibinfo{year}{1999}{\natexlab{b}}).

\bibitem[{\citenamefont{Mitroy}(2002)}]{mitroy02f}
\bibinfo{author}{\bibfnamefont{J.}~\bibnamefont{Mitroy}},
  \bibinfo{journal}{Phys.~Rev.~A} \textbf{\bibinfo{volume}{66}},
  \bibinfo{pages}{010501(R)} (\bibinfo{year}{2002}).

\bibitem[{\citenamefont{Drachman}(1979)}]{drachman79b}
\bibinfo{author}{\bibfnamefont{R.~J.} \bibnamefont{Drachman}},
  \bibinfo{journal}{Phys.~Rev.~A} \textbf{\bibinfo{volume}{19}},
  \bibinfo{pages}{1900} (\bibinfo{year}{1979}).

\bibitem[{\citenamefont{{Sullivan} et~al.}(2008)\citenamefont{{Sullivan},
  {Jones}, {Caradonna}, {Makochekanwa}, and {Buckman}}}]{sullivan08a}
\bibinfo{author}{\bibfnamefont{J.~P.} \bibnamefont{{Sullivan}}},
  \bibinfo{author}{\bibfnamefont{A.}~\bibnamefont{{Jones}}},
  \bibinfo{author}{\bibfnamefont{P.}~\bibnamefont{{Caradonna}}},
  \bibinfo{author}{\bibfnamefont{C.}~\bibnamefont{{Makochekanwa}}},
  \bibnamefont{and} \bibinfo{author}{\bibfnamefont{S.~J.}
  \bibnamefont{{Buckman}}}, \bibinfo{journal}{Rev.~Sci.~Instrum.}
  \textbf{\bibinfo{volume}{79}}, \bibinfo{pages}{113105}
  (\bibinfo{year}{2008}).

\bibitem[{\citenamefont{{Bylicki} and {Nicolaides}}(1995)}]{bylicki95a}
\bibinfo{author}{\bibfnamefont{M.}~\bibnamefont{{Bylicki}}} \bibnamefont{and}
  \bibinfo{author}{\bibfnamefont{C.~A.} \bibnamefont{{Nicolaides}}},
  \bibinfo{journal}{\pra} \textbf{\bibinfo{volume}{51}}, \bibinfo{pages}{204}
  (\bibinfo{year}{1995}).

\bibitem[{\citenamefont{Ho}(1983)}]{ho83a}
\bibinfo{author}{\bibfnamefont{Y.~K.} \bibnamefont{Ho}},
  \bibinfo{journal}{Phys.~Rep.} \textbf{\bibinfo{volume}{99}},
  \bibinfo{pages}{1} (\bibinfo{year}{1983}).

\bibitem[{\citenamefont{Zhou and Lin}(1995)}]{zhou95b}
\bibinfo{author}{\bibfnamefont{Y.}~\bibnamefont{Zhou}} \bibnamefont{and}
  \bibinfo{author}{\bibfnamefont{C.~D.} \bibnamefont{Lin}},
  \bibinfo{journal}{J.~Phys.~B} \textbf{\bibinfo{volume}{28}},
  \bibinfo{pages}{4907} (\bibinfo{year}{1995}).

\bibitem[{\citenamefont{{Han} et~al.}(2008)\citenamefont{{Han}, {Zhong},
  {Zhang}, and {Shi}}}]{han08b}
\bibinfo{author}{\bibfnamefont{H.}~\bibnamefont{{Han}}},
  \bibinfo{author}{\bibfnamefont{Z.}~\bibnamefont{{Zhong}}},
  \bibinfo{author}{\bibfnamefont{X.}~\bibnamefont{{Zhang}}}, \bibnamefont{and}
  \bibinfo{author}{\bibfnamefont{T.}~\bibnamefont{{Shi}}},
  \bibinfo{journal}{Phys.~Rev.~A} \textbf{\bibinfo{volume}{77}},
  \bibinfo{pages}{012721} (\bibinfo{year}{2008}).

\end{thebibliography}

\end{document}